\documentclass[twocolumn,showpacs,superscriptaddress,amsmath,amssymb,nofootinbib,secnumarabic,prd]{revtex4-2}
\usepackage{graphicx}
\usepackage{subfigure}
\usepackage{units}
\usepackage{multirow}
\usepackage{epsfig}
\usepackage{dcolumn}
\usepackage{booktabs}
\usepackage{appendix}
\usepackage{bm}
\usepackage{overpic}
\usepackage{color}
\usepackage{threeparttable}
\usepackage{lineno}
\usepackage{float}
\usepackage{epstopdf}
\usepackage[bookmarksnumbered, pdfstartview=FitH,colorlinks,urlcolor=blue, citecolor=blue,linkcolor=blue]{hyperref}
\usepackage[compat=1.1.0]{tikz-feynhand}
\newcommand{\gevcc}{\,\unit{GeV}/c^2}
\newcommand{\lambdacm}{\overline{\Lambda}{}_{c}^{-}}
\newcommand{\lambdacp}{\Lambda_{c}^{+}}
\newcommand{\modea}{pK_{S}^{0}}
\newcommand{\modeb}{pK^{-}\pi^+}
\newcommand{\modec}{pK_{S}^{0}\pi^0}
\newcommand{\moded}{pK_{S}^{0}\pi^+\pi^-}
\newcommand{\modee}{pK^{-}\pi^+\pi^0}
\newcommand{\modef}{p\pi^+\pi^-}
\newcommand{\modeaa}{\Lambda\pi^+}
\newcommand{\modebb}{\Lambda\pi^+\pi^0}
\newcommand{\modedd}{\Lambda\pi^+\pi^-\pi^+}
\newcommand{\modeaaa}{\Sigma^{0}\pi^+}
\newcommand{\modeccc}{\Sigma^{+}\pi^0}
\newcommand{\modeddd}{\Sigma^{+}\pi^+\pi^-}

\newcommand{\Modea}{\bar{p}K_{S}^{0}}
\newcommand{\Modeb}{\bar{p}K^{+}\pi^-}
\newcommand{\Modec}{\bar{p}K_{S}^{0}\pi^0}
\newcommand{\Moded}{\bar{p}K_{S}^{0}\pi^-\pi^+}
\newcommand{\Modee}{\bar{p}K^{+}\pi^-\pi^0}
\newcommand{\Modef}{\bar{p}\pi^-\pi^+}
\newcommand{\Modeaa}{\overline{\Lambda}{}\pi^-}
\newcommand{\Modebb}{\overline{\Lambda}{}\pi^-\pi^0}
\newcommand{\Modedd}{\overline{\Lambda}{}\pi^-\pi^+\pi^-}
\newcommand{\Modeaaa}{\overline{\Sigma}{}^{0}\pi^-}
\newcommand{\Modeccc}{\overline{\Sigma}{}^{-}\pi^0}
\newcommand{\Modeddd}{\overline{\Sigma}{}^{-}\pi^-\pi^+}

\newcommand{\sgmppimev}{\Sigma^+\pi^- e^+ \nu_e}
\newcommand{\sgmmpipev}{\Sigma^-\pi^+ e^+ \nu_e}
\newcommand{\sgmpTonpip}{\Sigma^+\to n\pi^+}
\newcommand{\sgmpToppi}{\Sigma^+\to p\pi^0}
\newcommand{\sgmmTonpim}{\Sigma^-\to n\pi^-}
\newcommand{\umiss}{U_{\rm miss}}
\newcommand{\Mmiss}{M^2_{\rm miss}}

\newcommand{\tabincell}[2]{\begin{tabular}{@{}#1@{}}#2\end{tabular}}

\newcommand{\BESIIIorcid}[1]{\href{https://orcid.org/#1}{\hspace*{0.1em}\raisebox{-0.45ex}{\includegraphics[width=1em]{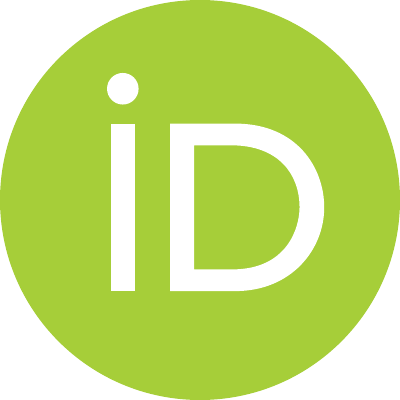}}}}

\begin{document}
\title{\bf\boldmath Evidence for the semileptonic decays \texorpdfstring{$\Lambda_c^{+} \to \Sigma^{\pm} \pi^{\mp} e^+ \nu_e$}{Lc->Sgmpiev}}
%\date{\it \small \bf \today}
\author{
%% Saved at => 2024-11-28
M.~Ablikim$^{1}$\BESIIIorcid{0000-0002-3935-619X},
M.~N.~Achasov$^{4,b}$\BESIIIorcid{0000-0002-9400-8622},
P.~Adlarson$^{76}$\BESIIIorcid{0000-0001-6280-3851},
X.~C.~Ai$^{81}$\BESIIIorcid{0000-0003-3856-2415},
R.~Aliberti$^{35}$\BESIIIorcid{0000-0003-3500-4012},
A.~Amoroso$^{75A,75C}$\BESIIIorcid{0000-0002-3095-8610},
Q.~An$^{72,58,\dagger}$,
Y.~Bai$^{57}$\BESIIIorcid{0000-0001-6593-5665},
O.~Bakina$^{36}$\BESIIIorcid{0009-0005-0719-7461},
Y.~Ban$^{46,g}$\BESIIIorcid{0000-0002-1912-0374}
% ...
\\
\vspace{0.2cm}
(BESIII Collaboration)\\
\vspace{0.2cm} {\it
$^{1}$ Institute of High Energy Physics, Beijing 100049, People's Republic of China\\
% ...
$^{4}$ Budker Institute of Nuclear Physics SB RAS (BINP), Novosibirsk 630090, Russia\\
% ...
$^{35}$ Johannes Gutenberg University of Mainz, Johann-Joachim-Becher-Weg 45, D-55099 Mainz, Germany\\
$^{36}$ Joint Institute for Nuclear Research, 141980 Dubna, Moscow region, Russia\\
% ...
$^{46}$ Peking University, Beijing 100871, People's Republic of China\\
% ...
$^{57}$ Southeast University, Nanjing 211100, People's Republic of China\\
$^{58}$ State Key Laboratory of Particle Detection and Electronics, Beijing 100049, Hefei 230026, People's Republic of China\\
% ...
$^{72}$ University of Science and Technology of China, Hefei 230026, People's Republic of China\\
% ...
$^{75}$ University of Turin and INFN, (A)University of Turin, I-10125, Turin, Italy; (B)University of Eastern Piedmont, I-15121, Alessandria, Italy; (C)INFN, I-10125, Turin, Italy\\
$^{76}$ Uppsala University, Box 516, SE-75120 Uppsala, Sweden\\
% ...
$^{81}$ Zhengzhou University, Zhengzhou 450001, People's Republic of China\\

\vspace{0.2cm}
$^{\dagger}$ Deceased\\
% ...
$^{b}$ Also at the Novosibirsk State University, Novosibirsk, 630090, Russia\\
% ...
$^{g}$ Also at State Key Laboratory of Nuclear Physics and Technology, Peking University, Beijing 100871, People's Republic of China\\
% ...
}
%% ends here %%

}
\begin{abstract}
Using $4.5\, \unit{fb^{-1}}$ of $e^+e^-$ collision data collected by the BESIII detector at center-of-mass energies between $4.600$ and $4.699\,\unit{GeV}$, we search for the semileptonic decays $\lambdacp \to \sgmppimev$ and $\lambdacp \to \sgmmpipev$ for the first time. Assuming their branching fractions are equal under isospin symmetry, evidence for $\lambdacp \to \Sigma^\pm\pi^\mp e^+\nu_e$ is reported with a significance of $3.6\sigma$. The corresponding branching fraction is measured to be $\mathcal{B}(\lambdacp \to \Sigma^\pm\pi^\mp e^+\nu_e) = (7.7^{+2.5}_{-2.3_{\rm 
 stat.}}\pm1.3_{\rm syst.})\times 10^{-4}$, which is consistent with quark model predictions within two standard deviations.
\end{abstract}

\maketitle
\section{Introduction}

The semileptonic~(SL) decays of the ground state charmed hadrons involve both the strong and weak interactions. 
For the lightest charmed baryon, $\Lambda_c^+$, these decays are of great interest for probing non-perturbative effects in Quantum Chromodynamics (QCD) in the charmed baryon system.
In recent years, significant progress has been achieved both experimentally and theoretically in the study of $\Lambda_c^+$ SL decays~\cite{Li:2021iwf,Cheng:2021vca,Cheng:2021qpd,Geng:2022fsr,Ke:2023qzc, Li:2025nzx}, which enrich our knowledge on the QCD mechanism in the charm sector. 

The dominant SL decay, $\Lambda_c^+ \to \Lambda l^+ \nu_l$ ($l=e$ or $\mu$), has been studied extensively by the BESIII collaboration. The branching fraction (BF) of $\Lambda_c^+ \to \Lambda e^+ \nu_e$ is measured to be $(3.56 \pm 0.13)\%$ and the $\Lambda_c^+ \to \Lambda$ form factors are also determined~\cite{BESIII:2022ysa, BESIII:2015ysy,BESIII:2023jxv, BESIII:2016ffj}. 
The four-body SL decay $\Lambda_c^+ \to p K^- e^+ \nu_e$ was first observed at BESIII with a BF of $(0.88\pm0.18)\times10^{-3}$~\cite{BESIII:2022qaf_pkev}. 
The BF of the inclusive SL decay $\Lambda_c^+ \to X e^+ \nu_e$ is measured to be $(4.06 \pm 0.13)\%$~\cite{BESIII:2022cmg_eX, BESIII:2018mug}, which agrees with the sum of the two exclusive BFs above but is 1.5 standard deviations lower. The SL decays $\Lambda_c^+ \to \Lambda \pi^+ \pi^- e^+ \nu_e$ and $\Lambda_c^+ \to p K^0_S \pi^- e^+ \nu_e$ have been searched for at BESIII and their BF upper limits at 90\% confidence level are $\mathcal{B}(\Lambda_c^+ \to \Lambda \pi^+ \pi^- e^+ \nu_e) < 3.9\times10^{-4}$ and $\mathcal{B}(\Lambda_c^+ \to p K^0_S \pi^- e^+ \nu_e) < 3.3\times10^{-4}$~\cite{BESIII:2023jem}.
In addition, the $\Lambda_c^+$ SL decays to $\Lambda$ excited states are investigated via the process $\Lambda_c^+ \to p K^- e^+ \nu_e$. Evidence for $\Lambda_c^+ \to \Lambda(1405) e^+ \nu_e$ and $\Lambda_c^+ \to \Lambda(1520) e^+ \nu_e$ are found with significances of $3.2\sigma$ and $3.3\sigma$, respectively~\cite{BESIII:2022qaf_pkev}. 
These results stimulate further exploration into other $\Lambda_c^+$ SL decays, such as the $\Sigma \pi$ final states from $\Lambda_c^+ \to \Lambda^* e^+\nu_e,\, \Lambda^* \to\Sigma \pi$.

Theoretically, the SL decay $\Lambda_c^+ \to \Lambda^* e^+ \nu_e$ is studied with semi-relativistic and non-relativistic constituent quark models~\cite{Pervin:2005ve, Hussain:2017lir}, the chiral unitary approach~\cite{Ikeno:2015xea} and Lattice QCD~\cite{Meinel:2021grq}. 
For the BF of $\Lambda_c^+ \to \Lambda(1520)e^+ \nu_e)$, predictions based on the quark models range from $0.06\%$ to $0.12\%$~\cite{Pervin:2005ve, Hussain:2017lir}, while Lattice QCD predicts \mbox{$(5.12\pm0.82)\times10^{-4}$}~\cite{Meinel:2021grq}.
The quark model predictions of \mbox{$\mathcal{B}(\Lambda_c^+ \to \Lambda(1405)e^+ \nu_e)$} range from $0.24\%$ to $0.59\%$~\cite{Pervin:2005ve, Hussain:2017lir}. However, the chiral unitary approach, assuming the $\Lambda(1405)$ is a molecular state, predicts \mbox{$\mathcal{B}(\Lambda_c^+ \to \Lambda(1405)e^+ \nu_e)$} to be $2\times10^{-5}$~\cite{Ikeno:2015xea}, which is nearly two orders of magnitude lower than quark model predictions.
Whether the $\Lambda(1405)$ is a three-quark bound state or a molecular state has been a long-standing mystery.
Since the $\Lambda(1405)$ decays into $\Sigma\pi$ with a large BF, the SL decay $\Lambda_c^+ \to \Sigma \pi e^+ \nu_e$ is a good channel to search for $\Lambda_c^+ \to \Lambda(1405) e^+ \nu_e$ and shed light on the nature of the $\Lambda(1405)$.
Ref.~\cite{Hussain:2017lir} calculates the BF of $\Lambda_c^+ \to \Lambda^* e^+ \nu_e \to \Sigma \pi e^+ \nu_e$ to be $0.36\%$ using a non-relativistic quark model. 

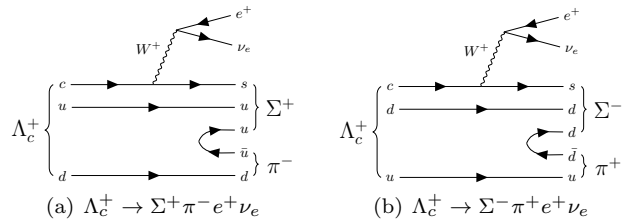
\begin{figure}[hpb]
    \centering
\subfigure[$\Lambda_c^+ \to \Sigma^+\pi^- e^+ \nu_e$]{
\resizebox{0.225\textwidth}{!}{
    \begin{tikzpicture}
        \begin{feynhand}
\vertex (w) at (2.5,3.2);
\vertex[particle] (e) at (4,3.6) {$e^+$};
\vertex[particle] (v) at (4,2.8) {$\nu_e$};
\vertex[particle] (a1) at (0,2) {$c$};
\vertex[particle] (a3) at (4,2) {$s$};
\vertex (a2) at (2,2);
\propag[fer] (a1) to (a2);
\propag[fer] (a2) to (a3);
\propag[bos] (a2) to (w);
\propag[fer] (e) to (w);
\propag[fer] (w) to (v);
\vertex[particle] (u1) at (0,1.5) {$u$};
\vertex[particle] (u2) at (4,1.5) {$u$};
\propag[fer] (u1) to (u2);
\vertex[particle] (d1) at (0,0) {$d$};
\vertex[particle] (d2) at (4,0) {$d$};
\propag[fer] (d1) to (d2);
\vertex (b) at (3,0.75);
\vertex[particle] (b1) at (4,1) {$u$};
\vertex[particle] (b2) at (4,0.5) {$\bar{u}$};
\propag[fer] (b2) to[out = 180, in = 290] (b);
\propag[fer] (b) to[out = 70, in = 180] (b1);
\node at (1.9, 2.8) {$W^+$};
        \end{feynhand}
        \draw [decorate,decoration={brace,amplitude=5pt},xshift=0pt,yshift=0pt]
        (-0.2,0) -- (-0.2,2) node [black,midway,xshift=-0.6cm] {\Large{$\Lambda_c^+$}};
        \draw [decorate,decoration={brace,amplitude=3pt},xshift=0pt,yshift=0pt]
        (4.2,2) -- (4.2,1) node [black,midway,xshift=0.6cm] {\Large{$\Sigma^+$}};
        \draw [decorate,decoration={brace,amplitude=3pt,mirror},xshift=0pt,yshift=0pt]
        (4.2,0) -- (4.2,0.5) node [black,midway,xshift=0.6cm] {\Large{$\pi^-$}};
    \end{tikzpicture}
}
}
\subfigure[$\Lambda_c^+ \to \Sigma^-\pi^+ e^+ \nu_e$]{
\resizebox{0.225\textwidth}{!}{
    \begin{tikzpicture}
        \begin{feynhand}
\vertex (w) at (2.5,3.2);
\vertex[particle] (e) at (4,3.6) {$e^+$};
\vertex[particle] (v) at (4,2.8) {$\nu_e$};
\vertex[particle] (a1) at (0,2) {$c$};
\vertex[particle] (a3) at (4,2) {$s$};
\vertex (a2) at (2,2);
\propag[fer] (a1) to (a2);
\propag[fer] (a2) to (a3);
\propag[bos] (a2) to (w);
\propag[fer] (e) to (w);
\propag[fer] (w) to (v);
\vertex[particle] (u1) at (0,1.5) {$d$};
\vertex[particle] (u2) at (4,1.5) {$d$};
\propag[fer] (u1) to (u2);
\vertex[particle] (d1) at (0,0) {$u$};
\vertex[particle] (d2) at (4,0) {$u$};
\propag[fer] (d1) to (d2);
\vertex (b) at (3,0.75);
\vertex[particle] (b1) at (4,1) {$d$};
\vertex[particle] (b2) at (4,0.5) {$\bar{d}$};
\propag[fer] (b2) to[out = 180, in = 290] (b);
\propag[fer] (b) to[out = 70, in = 180] (b1);
\node at (1.9, 2.8) {$W^+$};
        \end{feynhand}
        \draw [decorate,decoration={brace,amplitude=5pt},xshift=0pt,yshift=0pt]
        (-0.2,0) -- (-0.2,2) node [black,midway,xshift=-0.6cm] {\Large{$\Lambda_c^+$}};
        \draw [decorate,decoration={brace,amplitude=3pt},xshift=0pt,yshift=0pt]
        (4.2,2) -- (4.2,1) node [black,midway,xshift=0.6cm] {\Large{$\Sigma^-$}};
        \draw [decorate,decoration={brace,amplitude=3pt,mirror},xshift=0pt,yshift=0pt]
        (4.2,0) -- (4.2,0.5) node [black,midway,xshift=0.6cm] {\Large{$\pi^+$}};
    \end{tikzpicture}
    }
}
    \caption{Feynman diagrams of the $\lambdacp\to\Sigma^\pm\pi^\mp e^+\nu_e$ processes.}
    \label{fig:Feynman}
\end{figure}

In this paper, we search for the SL decays $\Lambda_c^+ \to \Sigma^\pm \pi^\mp e^+ \nu_e$, based on the $e^+e^-$ collision data with a total integrated luminosity of $4.5\ \unit{fb^{-1}}$ collected at the center-of-mass energies $\sqrt{s} = 4.600, 4.612, 4.628, 4.641, 4.661, 4.682$, and $4.699\ \unit{GeV}$~\cite{BESIII:2022ulv, BESIII:2015zbz, BESIII:2022dxl} by the BESIII detector. 
The decay Feynman diagrams are shown in Fig.~\ref{fig:Feynman} and the BFs of $\Lambda_c^+ \to \Sigma^\pm \pi^\mp e^+ \nu_e$ are expected to be the same under isospin symmetry. We assume the same BFs combining the two decay modes to improve the statistics.
Throughout this paper, charge conjugate channels are always implied.

\section{BESIII detector and Monte Carlo samples}
The BESIII detector, a cylindrical spectrometer on the BEPCII electron-positron collider~\cite{Yu:2016cof}, consists of a helium-based multilayer drift chamber (MDC), a plastic scintillator time-of-flight system (TOF), a CsI(Tl) electromagnetic calorimeter (EMC) inside a superconducting solenoidal magnet providing a 1.0 T magnetic field, and a muon counter. 
The design and performance of the BESIII detector is described in detail in 
Refs.~\cite{BESIII:2009fln_bes3, BESIII:2020nme}.

Monte Carlo (MC) samples are generated based on {\sc geant4}~\cite{GEANT4:2002zbu,Huang:2022wuo} to estimate the efficiency of the reconstruction and analyze the background. 
To analyze the potential background, an inclusive MC sample is generated simulating the inclusive $\lambdacp\lambdacm$ decays and other possible processes in the $e^+e^-$ annihilation, including open charm mesons, initial state radiation (ISR) produced lower charmonium states, and continuum $q\bar{q}$ process. 
The beam energy spread and ISR effects are incorporated with {\sc kkmc}~\cite{Jadach:2000ir}. The final state radiation is simulated with {\sc photos}~\cite{Richter-Was:1992hxq_photos}. 
The decays of the particles are modeled by {\sc evtgen}~\cite{Lange:2001uf_EvtGen, Ping:2008zz_EvtGen}, with the known decay BFs input from Particle Data Group (PDG)~\cite{pdg2022}.  
The signal MC samples simulate $e^+e^-\to\lambdacp\lambdacm$, where $\lambdacm$ decays to twelve hadronic modes described below and $\lambdacp$ decays to the signal processes $\lambdacp\to\Sigma^\pm\pi^\mp e^+ \nu_e$. The signal processes are generated uniformly in phase space.

\section{Analysis}
\subsection{Analysis method}
The data sets used are collected near the $\lambdacp\lambdacm$ pair production threshold, so that the $\lambdacp\lambdacm$ pairs are produced without additional hadrons. The double-tag (DT) technique, firstly introduced by the MARK-III collaboration~\cite{MARK-III:1985hbd}, is suitable in this situation to reduce backgrounds and infer the four-momentum of the unobserved neutrino. Twelve hadronic decay modes, $\Modea$, $\Modeb$, $\Modec$, $\Moded$, $\Modee$, $\Modeaa$, $\Modebb$, $\Modedd$, $\Modeaaa$, $\Modeccc$, $\Modeddd$ and $\Modef$ are used to fully reconstruct a $\lambdacm$, which is called a single tag (ST) $\lambdacm$. 
Then, the signal processes are searched for in the particles recoiling against the reconstructed $\lambdacm$. An event with both a ST $\lambdacm$ and a signal $\lambdacp$ reconstructed is called a DT event. 

Since neutrinos cannot be detected by the BESIII detector, kinematic variables constructed with the missing energy $E_{\rm miss}$ and missing momentum $\vec{p}_{\rm miss}$ are used to identify the presence of a neutrino. 
They are $E_{\mathrm{miss}} = E_{\mathrm{beam}} - E_{\mathrm{SL}}$ and \mbox{$\vec{p}_{\mathrm{miss}} = \vec{p}_{\lambdacp} - \vec{p}_{\mathrm{SL}}$} in the initial $e^+e^-$ rest frame, where $E_{\mathrm{SL}}$ and $\vec{p}_{\mathrm{SL}}$ are the energy and momentum of the reconstructed $\Sigma\pi e^+$ combination in the SL decay and $E_{\rm beam}$ is the beam energy. 
The $\lambdacp$ momentum $\vec{p}_{\lambdacp}$ is determined by \mbox{$\vec{p}_{\lambdacp} = -\hat{p}_{\mathrm{tag}}\sqrt{E^{2}_{\mathrm{beam}}/c^{2}-m^{2}_{\lambdacm}c^{2}}$}, where $\hat{p}_{\mathrm{tag}}$ is the direction of the momentum of the ST $\lambdacm$, and $m_{\lambdacm}$ is the known $\lambdacm$ mass~\cite{pdg2022}. 
The values of 
\begin{eqnarray}
U_{\mathrm{miss}} = E_{\mathrm{miss}} - c|{\vec{p}_{\mathrm{miss}}}|
\end{eqnarray}
and 
\begin{eqnarray}
M_{\rm miss}^2 = E^2_{\rm miss}/c^{4}-|\vec{p}_{\rm miss}|^2/c^{2}
\end{eqnarray}
are expected to peak around zero in the signal SL process. %given the near zero mass of neutrino. 
The $\umiss$ and $\Mmiss$ variables are used to fit for the total DT yield combining data samples of all ST modes and energy points. 

The BFs of the signal decays are calculated by 
\begin{eqnarray}
\mathcal{B}_{s} = \frac{N^{\rm DT}}{\mathcal{B}_{\rm inter} \, N^{\rm ST} \, \varepsilon_{s}},
\label{eq:bf}
\end{eqnarray}
where $N^{\rm DT}$ is the total DT yield and $\mathcal{B}_{\rm inter}$ is the product BF of the intermediate decays in the signal process.
The averaged efficiency of detecting the signal process, $\varepsilon_{s}$, is defined as
\begin{eqnarray}
\varepsilon_{s} \,=\, \frac{1}{N^{\rm ST}} \, \sum_{i,j} \left (\frac{N_{ij}^{\rm ST}}{\varepsilon_{ij}^{\rm ST}} \, \varepsilon_{ij}^{\rm DT} \right), \label{eq:effective_eff}
\end{eqnarray}
where $\varepsilon_{ij}^{\rm ST}$, $\varepsilon_{ij}^{\rm DT}$ and $N_{ij}^{\rm ST}$ are the ST efficiency, DT efficiency and ST yield of the ST mode $i$ at the $j$-th energy point, respectively, and $N^{\rm ST}$ is the total ST yield $N^{\rm ST} = \sum_{i,j}N_{ij}^{\rm ST}$.

\subsection{Single tag event selection}
\begin{figure}[htb!]
\centering
\includegraphics[width=0.99\linewidth]{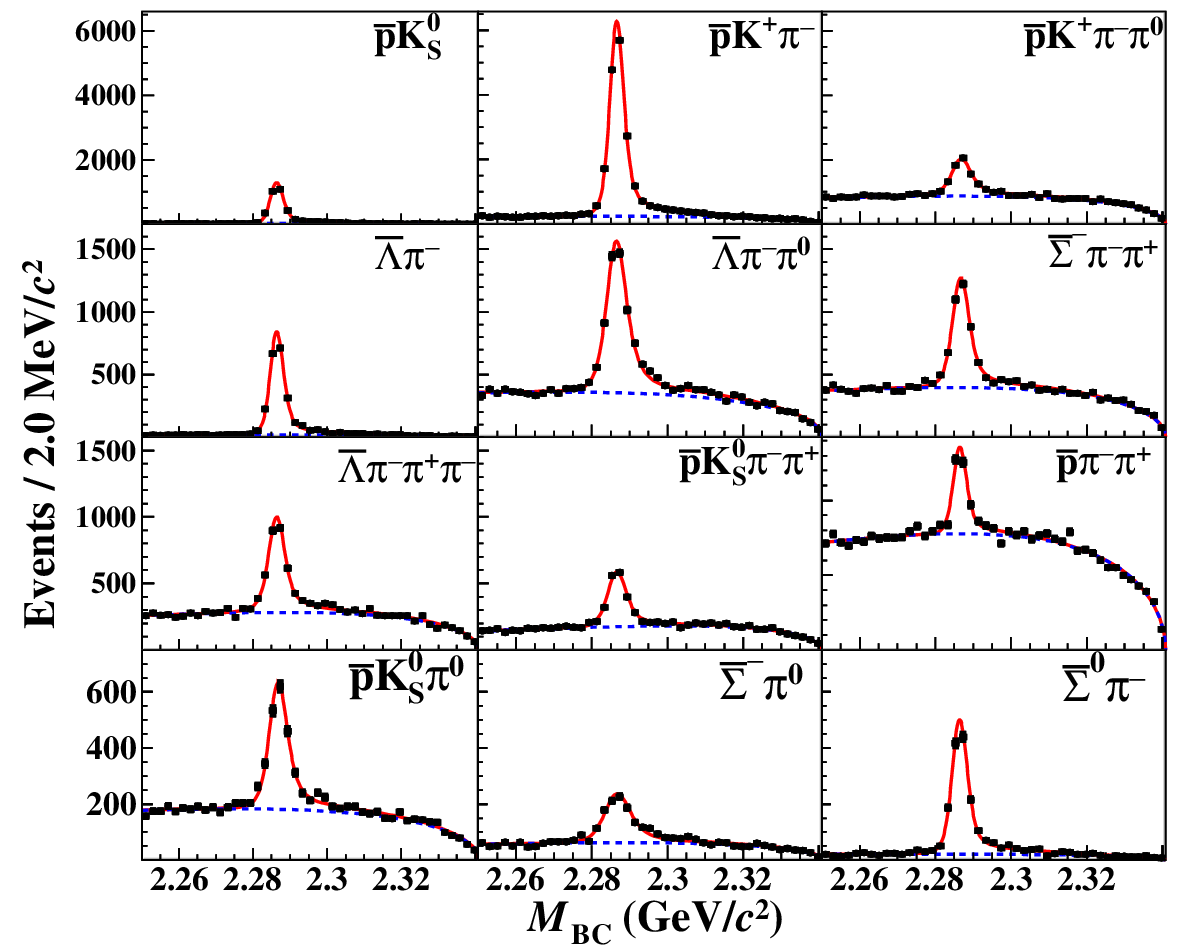}
\caption{The $M_{\rm BC}$ distributions in the twelve ST modes of the data sample at $\sqrt{s} = 4.682\,\unit{GeV}$. The black points with error bars are data, the red curves are the total fit results, and the dashed blue curves show the fitted backgrounds.
}
\label{fig:ST}
\end{figure}

We select ST $\lambdacm$ candidates following the selection criteria used in the previous BESIII analysis~\cite{BESIII:2022xne_ST}, except for removing the requirement on the ratio of the decay length of $\overline{\Lambda}{}$ to its uncertainty in order to increase the efficiency.
If multiple ST $\lambdacm$ candidates per tag mode per charge per event satisfy the selection criteria, the candidate with the minimum energy difference, $|\Delta E| = |E - E_{\rm beam}|$, is retained, where $E$ is the energy of the $\lambdacm$ candidate in the initial $e^+e^-$ rest frame.
Subsequently, the ST $\lambdacm$ candidates are identified within the beam constrained mass range $M_{\rm BC}\in(2.28, 2.30)\gevcc$, where 
\begin{eqnarray}
M_{\rm BC} = \sqrt{E_{\rm beam}^2/c^4 - |\vec{p}\,|^2/c^2}. 
\end{eqnarray}
Here, $\vec{p}$ is the momentum of the ST $\lambdacm$ candidate in the initial $e^+e^-$ rest frame. 
The ST yields and efficiencies are obtained by the fitting methods described in Ref.~\cite{BESIII:2022xne_ST}. 
The $M_{\rm BC}$ distributions in the twelve ST modes in the data sample at $\sqrt{s}=4.682\,\unit{GeV}$ are shown in Fig.~\ref{fig:ST}. 
The ST yields and efficiencies are listed in Table~\ref{tab:ST}. 
The yields are summed over all energy points $N^{\rm ST}_i = \sum_{j}N^{\rm ST}_{ij}$ and the efficiencies are averaged according to the yields of each energy point $\varepsilon^{\rm ST}_i = \sum_{j}N^{\rm ST}_{ij}/\sum_{j}\frac{N^{\rm ST}_{ij}}{\varepsilon^{\rm ST}_{ij}}$ .
The sum of ST yields is $N^{\rm ST}=120350\pm464$, where the uncertainty is statistical only. 

\begin{table}[hptb]
  \begin{center}
     \caption{Total yield, $N^{\rm ST}_i$, and averaged efficiencies, $\varepsilon^{\rm ST}_i$, in each ST mode. The ST yields are summed over all energy points and the ST efficiencies are averaged according to the yields at different energy points. The uncertainties are statistical only. The BFs of intermediate decays are not included in the efficiencies.}
  \begin{tabular}{lll}
    \hline \hline
    Tag mode & $N^{\rm ST}_i$ & $\varepsilon^{\rm ST}_i (\%)$ \\ \hline
$\modea$ & $8839\pm98$ & $49.45\pm0.08$ \\
$\modeb$ & $46530\pm236$ & $46.12\pm0.05$ \\
$\modec$ & $4422\pm100$ & $19.33\pm0.07$ \\
$\moded$ & $4052\pm91$ & $20.06\pm0.07$ \\
$\modee$ & $12481\pm191$ & $17.15\pm0.05$ \\
$\modeaa$ & $5850\pm79$ & $45.15\pm0.12$ \\
$\modebb$ & $13270\pm162$ & $20.57\pm0.05$ \\
$\modedd$ & $6607\pm125$ & $16.54\pm0.05$ \\
$\modeaaa$ & $3859\pm70$ & $27.69\pm0.08$ \\
$\modeccc$ & $2216\pm66$ & $21.18\pm0.10$ \\
$\modeddd$ & $8028\pm138$ & $22.14\pm0.05$ \\
$\modef$ & $4196\pm135$ & $55.97\pm0.17$ \\
	\hline \hline
   \end{tabular}
   \label{tab:ST}
  \end{center}
  \end{table}

\subsection{Signal event selection}
The SL decay signals $\lambdacp \to \Sigma^{\pm} \pi^{\mp} e^+ \nu_e$ are reconstructed among the remaining tracks and showers recoiling against the reconstructed ST $\lambdacm$. 
The $\Sigma^+$ candidates are reconstructed with two dominant decays, $\Sigma^+\to p\pi^0$ and $\Sigma^+\to n\pi^+$, while the $\Sigma^-$ candidates are reconstructed with the dominant $\Sigma^-\to n\pi^-$ decay.  

Charged tracks are required to have a polar angle $\theta$, defined as the angle between the charged track and the $z$ axis (the symmetry axis of the MDC), satisfying $\vert\!\cos\theta\vert<0.93$, and have a distance of closest approach to interaction point along the $z$ axis of less than 20\,cm. 
The number of charged tracks is required to be exactly three. 
For the candidate tracks of the positron and pion from the $\lambdacp$ decay, the distance of closest approach to interaction point must be less than 1\,cm perpendicular to the $z$ axis and less than 10\,cm along the $z$ axis. 
Particle identification is performed using the time of flight measured by the TOF and the ${\rm d}E/{\rm d} x$ measured by the MDC. Likelihoods $\mathcal{L}(h)$ are calculated with different hypotheses, $(h = (p, K, \pi)$. 
The charged tracks are identified as protons when $\mathcal{L}(p)>0$, $\mathcal{L}(p) > \mathcal{L}(\pi)$ and $\mathcal{L}(p) > \mathcal{L}(K)$, and identified as pions when $\mathcal{L}(\pi)>0$, $\mathcal{L}(\pi) > \mathcal{L}(K)$. 
The information from the EMC is added for the identification of positron candidates. Tracks with $\mathcal{L}(e)>0$ and $\mathcal{L}(e)/(\mathcal{L}(e) + \mathcal{L}(\pi) + \mathcal{L}(K))>0.999$ are identified as positrons.

Photon candidates are identified using showers in the EMC. The energy deposit of each shower is required to be larger than 25\,MeV for showers in the barrel region ($\vert\!\cos\theta\vert<0.80$), or larger than 50\,MeV in the endcap region ($0.86<\vert\!\cos\theta\vert<0.92$). 
The difference between the EMC time and the event start time is required to be within $700\,\unit{ns}$ to suppress electronic noise and showers unrelated to the collision events.
The $\pi^0$ candidates are reconstructed via their diphoton decay and the invariant mass of the photon pair $M_{\gamma\gamma}$ is required to be within $(0.115,\,0.150)\gevcc$. A kinematic fit constraining $M_{\gamma\gamma}$ to the known $\pi^0$ mass~\cite{pdg2022} (1C) is performed and the $\chi^2_{\rm 1C}$ is required to be less than 200. The fitted four-momenta of $\pi^0$ candidates are used in the further analysis. 

The $\Sigma^+$ reconstructed by the decay $\Sigma^+\to p\pi^0$ is required to have the $p\pi^0$ invariant mass $M_{p\pi^0}$ in the range $(1.176,\,1.200)\gevcc$. A similar kinematic fit constraining $M_{p\pi^0}$ to the known $\Sigma^+$ mass $m_{\Sigma^+}$~\cite{pdg2022} is performed and the fitted four-momentum of $\Sigma^+$ is used for the further analysis. 
If multiple candidates are selected, the one with the minimum mass difference $|M_{p\pi^0}-m_{\Sigma^+}|$ is retained. 

\begin{figure*}[htb!]
\centering
{\includegraphics[width=0.325\linewidth]{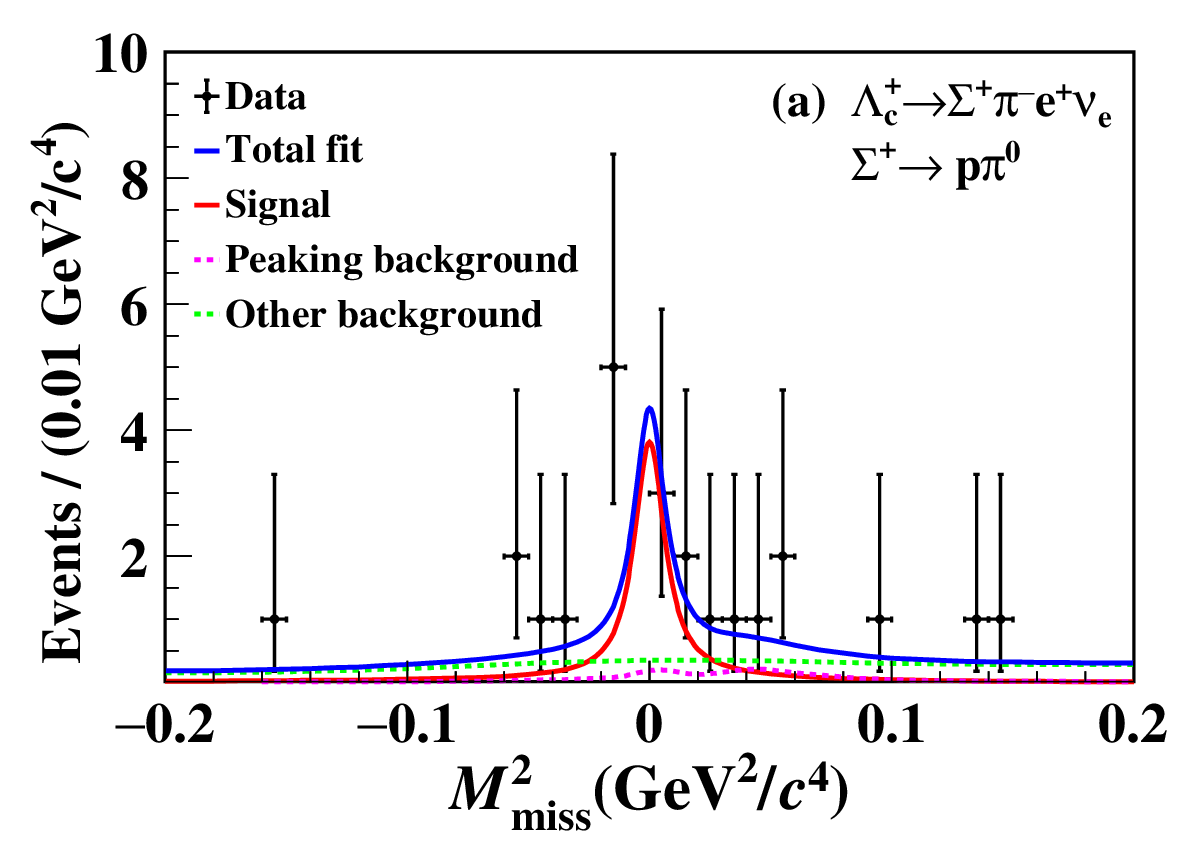}}
{\includegraphics[width=0.325\linewidth]{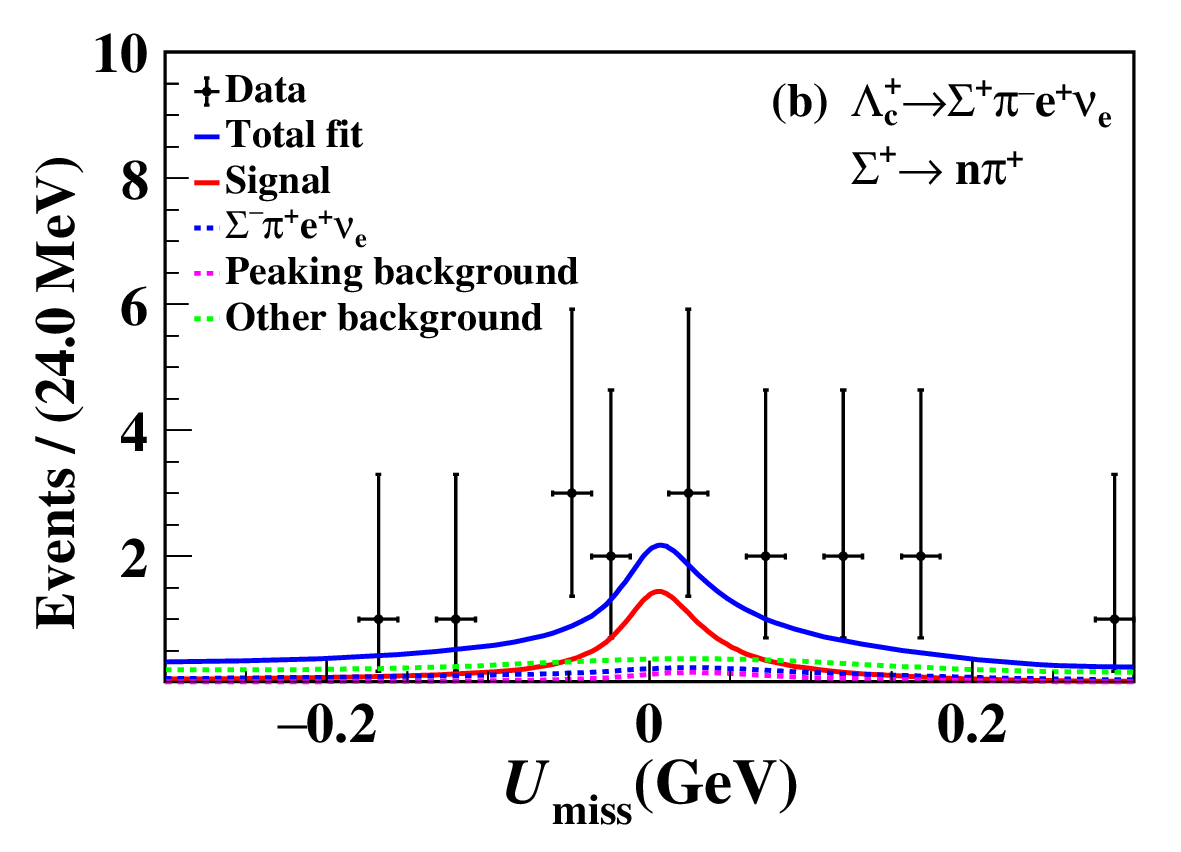}}
{\includegraphics[width=0.325\linewidth]{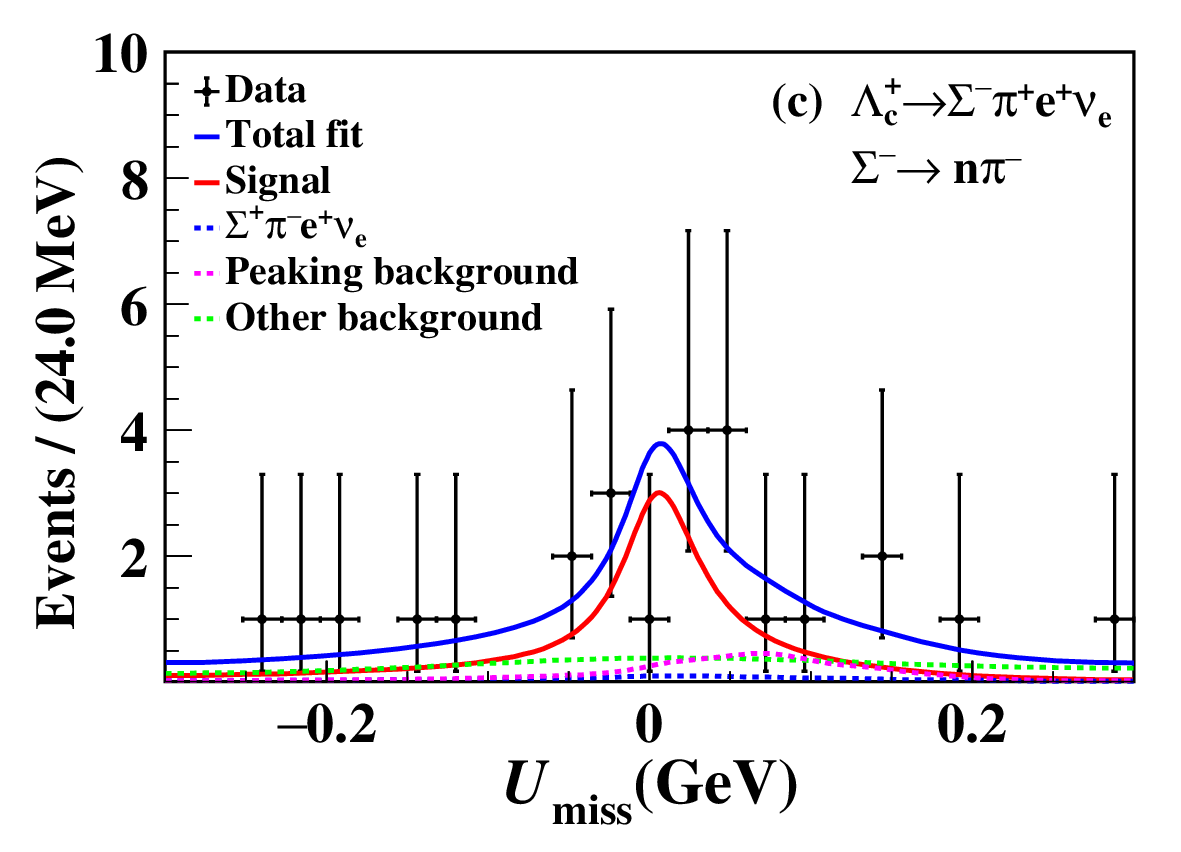}}
%\label{fig:sgmp_p}
%\label{fig:sgmp_n}
%\label{fig:sgmm_n}
\caption{\label{fig:sgm}
(a) The $M_{\rm{miss}}^2$ distributions of $\lambdacp \to \sgmppimev,\, \sgmpToppi$ candidates. 
(b) The $\umiss$ distributions of $\lambdacp \to \sgmppimev,\, \sgmpTonpip$ candidates. 
(c) The $\umiss$ distributions of $\lambdacp \to \sgmmpipev,\, \sgmmTonpim$ candidates. 
The black dots with error bars represent data and blue lines  represent the fit results. 
The peaking backgrounds, represented by the magenta dashed lines, consists of $\Sigma^+\pi^+\pi^-$ and $\Sigma^+\omega$ in (a)(b) and $\Sigma^-\pi^+\pi^+$ and $\Sigma^-\pi^+\pi^+\pi^0$ in (c).
}
\label{fig:fit}
\end{figure*}

Based on a study of the inclusive MC sample, we find that the main backgrounds in the signal candidates $\lambdacp \to \sgmppimev$, $\Sigma^+\to p \pi^0$ are from $\lambdacp \to \Sigma^+ \pi^+\pi^-$, $\Lambda e^+ \nu_e$, and $\Sigma^+ \omega$. Hence, the invariant mass of $p\pi^-$ in the final state is required to be larger than $1.13\gevcc$, to suppress the $\lambdacp \to \Lambda e^+ \nu_e, \Lambda \to p\pi^-$ background. 
Background events from $\lambdacp \to \Sigma^+ \pi^+\pi^-$ are reduced by requiring $M_{\Sigma^+\pi^-\pi(e)^+}<2.27\gevcc$, where $M_{\Sigma^+\pi^-\pi(e)^+}$ is the invariant mass of the $\Sigma^+\pi^-\pi^+$ system obtained by replacing the positron mass with the $\pi^+$ mass. 
An additional source of backgrounds arises from $\gamma$-conversions in $\lambdacp$ inclusive decays containing a $\pi^0$. These backgrounds are suppressed by requiring $\cos\theta_{\pi,e}<0.95$, where $\theta_{\pi,e}$ is the angle between the momenta of the $\pi^-$ and $e^+$ candidates. 
The resulting $M_{\rm miss}^2$ distribution of the accepted candidates in data is shown in Fig.~\ref{fig:sgm}(a). A cluster corresponding to the signal is observed near zero.  

The signal processes $\lambdacp\to\sgmppimev$, $\sgmpTonpip$ and $\lambdacp\to\sgmmpipev$, $\sgmmTonpim$ are jointly analyzed, as they share the same final states.
The charged pions and positrons are identified with the same criteria as those used for the $\Sigma^+\to p \pi^0$ mode. For the neutron final states, only shower information from the EMC is available; this provides the polar and azimuthal angles of the candidate neutron showers.  These showers must pass the same criteria used for the photon candates.  We then apply a constraint of $U_{\rm miss}=0$ to determine the magnitude of neutron momentum.  Among the shower candidates that provide a physical solution for this magnitude, the one with the largest deposited energy is  identified as the neutron-induced shower.  
To select the $\Sigma$ candidates, the invariant masses $M_{n\pi^+}$ and $M_{n\pi^-}$ are calculated, where $M_{n\pi^+}\in(1.15,\,1.23)\gevcc$ and $M_{n\pi^-}\in(1.16,\,1.24)\gevcc$ are required, respectively.
In order to extract the signal, we perform a kinematic fit constraining the invariant mass of $n\pi^+$($n\pi^-$) to the known $\Sigma^+$($\Sigma^-$) mass to derive the magnitude of neutron momentum again, independent of the $\umiss = 0$ hypothesis.

For these neutron modes, similar requirements, $M_{\Sigma^+\pi^-\pi(e)^+}<2.27\gevcc$ and $\cos\theta_{\pi,e}<0.95$ are adopted to remove backgrounds similar to those in the $\Sigma^+\to p\pi^0$ mode. 
Additionally, contamination from $\pi^0$-induced showers needs to be controlled.
We suppress such backgrounds by vetoing events with detected $\pi^0$ candidates, according to the following criteria. 
The showers considered are required to have a ratio of deposited energy in $3\times3$ crystal cells over that in $5\times5$ crystal cells, $E_{3\times3}/E_{5\times5}$, larger than 0.9 and a crystal energy-deposit second moment smaller than $20\,\unit{cm^2}$, in order to distinguish the showers from photons and neutrons, and especially anti-neutrons.
The $\pi^0$ candidates are required to have $M_{\gamma\gamma} \in (0.115,\,0.150)\gevcc$ and $\chi^2_{\rm 1C}<20$. 
After this veto, the $U_{\rm miss}$ distributions of the remaining candidates in data are shown in Fig.~\ref{fig:sgm}(b) and Fig.~\ref{fig:sgm}(c). %The signals are expected to peak near zero.

\subsection{Branching fraction measurement}
The DT signal yields are determined by fitting the $\Mmiss$ distribution for $\lambdacp \to \sgmppimev,\, \sgmpToppi$ and the $\umiss$ distributions for $\lambdacp \to \sgmppimev,\, \sgmpTonpip$ and $\lambdacp \to \sgmmpipev,\, \sgmmTonpim$.  The $\Mmiss$ distribution is used in the first case since it has better resolution, while the $\umiss$ distributions provide better separation from the significant peaking backgrounds in the modes involving neutron. 
A simultaneous unbinned maximum likelihood fit to the $\Mmiss$ and $\umiss$ distributions is performed by sharing the same BF for $\lambdacp \to \sgmppimev$ and $\lambdacp \to \sgmmpipev$ under the isospin symmetry assumption. 
In the fit, the signal shapes are extracted from the signal MC samples. The peaking backgrounds of \mbox{$\lambdacp \to \Sigma^+\pi^+\pi^-$} and $\lambdacp \to \Sigma^+\omega$ for the $\lambdacp \to \sgmppimev$ mode are depicted with high-statistics dedicated MC samples, and same treatment is applied to the peaking backgrounds of $\lambdacp \to \Sigma^-\pi^+\pi^+$ and $\lambdacp \to \Sigma^-\pi^+\pi^+\pi^0$ for the $\lambdacp \to \sgmmpipev$ mode.
The yields of the peaking backgrounds are fixed according to the known decay BFs~\cite{pdg2022}. 
In the modes involving neutrons, there is cross feed between $\lambdacp \to \sgmmpipev,\, \sgmmTonpim$ and $\lambdacp \to \sgmppimev,\, \sgmpTonpip$; the rates are estimated from the corresponding signal MC samples to be around 3\%. These cross feed backgrounds are described by MC extracted shapes, with their yields constrained by assuming the same BF for $\lambdacp \to \sgmppimev$ and $\lambdacp \to \sgmmpipev$.
The shapes of remaining backgrounds from $\lambdacp$ decays and continuum process are extracted according to the inclusive MC sample.

%The figures in the left column of 
Fig.~\ref{fig:fit} shows the simultaneous fit results that give $\mathcal{B}(\lambdacp \to \Sigma^{\pm} \pi^{\mp} e^+ \nu_e) = (7.7^{+2.5}_{-2.3})\times 10^{-4}$ with a $3.9\sigma$ statistical significance. The statistical significance is estimated using the test statistic $\sqrt{-2 \,\Delta\ln\mathcal{L}}$, where $\Delta\ln\mathcal{L}$ is the change of $\ln\mathcal{L}$ in the likelihood fit with and without including the signal. Here, the systematic uncertainties are not taken into account in the statistical significance.
For the $\lambdacp \to \sgmppimev$ mode, the signal yields are estimated to be $9.6^{+3.2}_{-2.9}$ in the mode $\sgmpToppi$ and $6.4^{+2.1}_{-1.9}$ in the mode $\sgmpTonpip$, while for the $\lambdacp \to \sgmmpipev$, $\sgmmTonpim$ mode, the signal yield is $13.3^{+4.4}_{-4.0}$.
The averaged reconstruction efficiencies, $\varepsilon_s$, are $(20.3\pm0.1)\%$ for $\lambdacp\to\sgmppimev,\, \sgmpToppi$ and $(14.3\pm0.1)\%$ for $\lambdacp \to \sgmppimev,\,\sgmpTonpip$ and $\lambdacp \to \sgmmpipev$, $\sgmmTonpim$. All the uncertainties are statistical only.

Separate fits for the \mbox{$\lambdacp \to \sgmppimev$} and $\sgmmpipev$ decays are also performed, where the \mbox{$\sgmpToppi$} and $\sgmpTonpip$ cases share the same BF of $\lambdacp \to \sgmppimev$. %The fit results are shown in the right column of Fig.~\ref{fig:fit}. 
This method results in \mbox{$\mathcal{B}(\lambdacp \to \sgmppimev) = (7.8^{+3.5}_{-3.0})\times10^{-4}$} and \mbox{$\mathcal{B}(\lambdacp \to \sgmmpipev) = (7.6^{+3.8}_{-3.5})\times10^{-4}$}, which are consistent with the symmetry.
Their statistical significances are $3.1\sigma$ for \mbox{$\lambdacp \to \sgmppimev$} and $2.4\sigma$ for $\lambdacp \to \sgmmpipev$.

\section{Systematic Uncertainty}
Using the DT method, the systematic uncertainties on the ST side are mostly canceled.
The remaining systematic uncertainties can be categorized into multiplicative and additive systematic uncertainties. 
The systematic uncertainties from the estimation of the signal processes efficiencies and ST yields contribute multiplicatively to the BFs calculations. 
The systematic uncertainties from the simultaneous fit procedure affect the significance of the fit results; these are additive systematic uncertainties. 

%\subsection{Multiplicative Systematic Uncertainties}
The multiplicative systematic uncertainties from different sources are listed in Table \ref{tab:sys_err}.

\begin{table}[htpb]
  \begin{center}
     \caption{The multiplicative systematic uncertainties, in percent, for the BF measurements.}
  \footnotesize
  \begin{tabular}{lccc}
    \hline \hline
    	Source &  \tabincell{c}{$\Sigma^+\pi^- e^+ \nu_e,$ \\ $\Sigma^+ \to p \pi^0$ }  &  \tabincell{c}{$\Sigma^+\pi^- e^+ \nu_e,$ \\ $\Sigma^+ \to n \pi^+$ }  & \tabincell{c}{$\Sigma^-\pi^+ e^+ \nu_e,$ \\ $\Sigma^- \to n \pi^-$}  \\ \hline
		Neutron reconstruction & - & 8.0 & 8.0 \\
		$\pi^0$ veto & - & 11.7 & 11.7 \\
		$M_{p\pi^-}>1.13\gevcc$ & 6.6 & - & - \\
		$M_{\Sigma\pi\pi(e)^+}<2.27\gevcc$ & 0.1 & 0.2 & 0.2 \\	
		$\cos\theta_{\pi,e}<0.95$ & 4.8 & 4.8 & 4.8 \\
		$e^+$ tracking, PID  & 10.4 & 10.4 & 10.4 \\ 
		ST $\lambdacm$ yield & 1.0 & 1.0 & 1.0 \\
		MC statistics & 0.4 & 0.4 & 0.4 \\
		Quoted BFs & 0.6 & 0.6 & - \\
		MC model & 16.7 & 1.3 & 3.9 \\ 
		$\pi^\pm$ tracking, PID & 0.3 & 0.8 & 0.9 \\
		Proton tracking, PID & 1.4 & - & - \\
		$\pi^0$ reconstruction & 3.1 & - & - \\
		\hline
		Total  & 21.6 & 18.3 & 18.7 \\
		%fit & 3.6 & 8.7 & 2.0 \\\hline
	\hline \hline
   \end{tabular}
   \label{tab:sys_err}
  \end{center}
  \end{table}

Due to imperfect simulation of neutron showers in the EMC, systematic uncertainties related to neutron reconstruction are investigated using control samples of $\lambdacp \to \Sigma^+\pi^+\pi^-,\, \sgmpTonpip$ and $\lambdacp \to \Sigma^-\pi^+\pi^+$, $ \sgmmTonpim$. 
The difference in neutron reconstruction efficiencies between data and MC simulation gives a systematic uncertainty of $8.0\%$. Based on the same control sample, the systematic uncertainty of the $\pi^0$ veto is estimated to be $11.7\%$. 
The systematic uncertainty related to the $\Sigma^+$ and $\Sigma^-$ mass windows after the neutron momentum reconstruction is studied with the control sample $\lambdacp \to \Sigma^-\pi^+\pi^+\pi^0$ with a missing $\pi^0$ corresponding to the missing neutrino in the signal process. This systematic uncertainty is found to be negligible. 

The systematic uncertainties related to the \mbox{$\lambdacp \to \sgmppimev$}, $\sgmpToppi$ reconstruction are studied with the control sample $\lambdacp \to \Sigma^+\pi^+\pi^-,\, \sgmpToppi$. 
By comparing the acceptance rates of the requirements between data and MC simulation, the systematic uncertainty due to the $M_{p\pi^-}>1.13\gevcc$ requirement is estimated to be $6.6\%$ and those due to the $\Sigma^+$ mass window and $\chi^2_{\rm 1C} < 200$ requirement in $\pi^0$ reconstruction are negligible. 

Some sources of systematic uncertainties are common for the three reconstruction modes. 
The systematic uncertainty associated with the $M_{\Sigma^+\pi^-\pi(e)^+}<2.27\gevcc$ requirement is estimated to be $0.1\%$ for $\lambdacp \to \sgmppimev,\, \sgmpToppi$ and $0.2\%$ for the modes with neutrons. This estimation is performed by smearing the $M_{\Sigma^+\pi^-\pi(e)^+}$ distribution in signal MC sample according to the resolution difference between data and MC simulation obtained in the control samples of $\lambdacp \to \Sigma^+\pi^+\pi^-$  and $\lambdacp \to \Sigma^-\pi^+\pi^+$. 
The systematic uncertainty associated with the $\cos\theta_{\pi,e}<0.95$ requirement is estimated to be $4.8\%$ using control sample $D^0 \to K^0_s \pi^- e^+ \nu_e$.
The systematic uncertainty arising from the ST $\lambdacp$ yield is assigned to be $1\%$~\cite{BESIII:2022ysa}. 
The systematic uncertainty due to limited statistics of the MC samples is $0.4\%$ for all three decay modes. 
The systematic uncertainty from the known intermediate BFs is $0.6\%$ for both $\lambdacp \to \sgmppimev,\, \sgmpToppi$ and $\lambdacp \to \sgmppimev,\, \sgmpTonpip$, quoted from the uncertainties of $\mathcal{B}(\sgmpToppi)$, $\mathcal{B}(\pi^0\to\gamma\gamma)$ and $\mathcal{B}(\sgmpTonpip)$ from the PDG~\cite{pdg2022}. The uncertainty from $\mathcal{B}(\sgmmTonpim)$ is negligible.
The systematic uncertainty from the signal MC model is estimated by changing the MC model from phase space to model with $\Lambda(1405)$ and $\Lambda(1520)$ resonances. The form factors of $\lambdacp \to \Lambda^*$ and the ratio of  $\Lambda(1405)$ and $\Lambda(1520)$ are quoted from Ref.~\cite{Hussain:2017lir}.
The relative differences of the efficiencies before and after altering the MC model are assigned as the systematic uncertainties, which are $16.7\%$, $1.3\%$ and $3.9\%$ for \mbox{$\lambdacp \to \sgmppimev$}, $\sgmpToppi$, $\lambdacp \to \sgmppimev$, $\sgmpTonpip$ and $\lambdacp \to \sgmmpipev,\, \sgmmTonpim$, respectively.

The systematic uncertainties related to the tracking, PID of $e^+$, $p$, $\pi^\pm$ and the reconstruction of $\pi^0$ are evaluated by re-weighting the MC samples according to the efficiency differences between data and MC simulation obtained with control samples. The control samples are chosen to be $e^+e^-\to e^+e^-\gamma$ for $e^+$ tracking and PID, $J/\psi \to p \bar{p} \pi^+\pi^-$ for proton tracking and PID, \mbox{$e^+e^-\to K^+ K^-\pi^+\pi^-$} and $e^+e^-\to \pi^+\pi^-\pi^+\pi^-$ for charged pion tracking and PID,  $e^+e^-\to K^+ K^-\pi^+\pi^-\pi^0$ for $\pi^0$ reconstruction. These estimated      systematic uncertainties are listed in Table \ref{tab:sys_err}.
 
The additive systematic uncertainties from the $\Mmiss$ and $\umiss$ fits are determined by altering the background descriptions in the fit. The fixed magnitudes of peaking backgrounds are varied 10,000 times around the fixed values with a Gaussian re-sampling method using the uncertainties of the estimated background yields. The standard deviations of the distributions of the fitted central values are taken as the systematic uncertainties. 
The modeling of the ``other'' background is changed to a flat shape with a yield constrained with a Poisson distribution, whose mean is the value estimated according to the $M_{\rm BC}$ sideband in data. The differences of fit results from the nominal results are taken as the systematic uncertainties. The total systematic uncertainties from the fit are the quadratic sum of each variation of the fit model, giving $5.6\%$ for the simultaneous fit of $\lambdacp \to \Sigma^\pm \pi^\mp e^+ \nu_e$.
In addition, the systematic uncertainties for separate fits to $\lambdacp \to \sgmppimev$ and  $\lambdacp \to \sgmmpipev$ are evaluated to be $10.7\%$ and $2.0\%$, respectively.

To incorporate the additive systematic uncertainties, the significance is re-evaluated by taking the minimum significance among the alternative fit models, which results in statistical significance of $2.7\sigma$, $2.3\sigma$ and $3.6\sigma$ for $\lambdacp \to \sgmppimev$, $\lambdacp \to \sgmmpipev$ and simultaneous fit of $\lambdacp \to \Sigma^\pm \pi^\mp e^+ \nu_e$, respectively.

The multiplicative systematic uncertainties as listed in Table \ref{tab:sys_err} are incorporated by adding Gaussian constraints in the likelihood construction, whose errors are the relative systematic uncertainties.
In the simultaneous fit, the correlation of systematic uncertainties among the three different reconstruction modes are taken into account. 
The systematic uncertainties from $\cos\theta_{\pi,e}<0.95$ requirement, $e^+$ tracking and PID, and ST $\lambdacm$ yield are considered to be 100\% correlated among the three modes, and total $11.5\%$ in quadrature. The systematic uncertainties from neutron reconstruction, $\pi^0$ veto, $M_{\Sigma\pi\pi(e)^+}<2.27\gevcc$ requirement, $\pi^{\pm}$ tracking and PID are considered to be 100\% correlated between the two neutron-involved signal modes, and are $14.2\%$ in total. The other systematic uncertainties are assumed to be independent among the three modes. 

\section{Results}

For the simultaneous fit, the total systematic uncertainties are calculated by combining the multiplicative systematic uncertainty and the additive systematic uncertainty, as $\sigma_{\rm sys.} = \sqrt{\sigma_{\rm Gaus}^2 - \sigma_{\rm stat}^2 + \sigma_{\rm add}^2}$. Here, $\sigma_{\text{stat}}$ and $\sigma_{\text{Gaus}}$ represent the statistical uncertainties before and after incorporating Gaussian constraints to account for multiplicative systematic uncertainties, while $\sigma_{\text{add}}$ denotes the additive systematic uncertainty.
Under the isospin symmetry assumption, we obtain $\mathcal{B}(\lambdacp \to \Sigma^\pm\pi^\mp e^+\nu_e) = (7.7^{+2.5}_{-2.3}\pm1.3)\times 10^{-4}$. 
The separate BF results are $\mathcal{B}(\lambdacp \to \sgmppimev) = (7.8^{+3.5}_{-3.0}\pm1.6)\times10^{-4}$ and $\mathcal{B}(\lambdacp \to \sgmmpipev) = (7.6^{+3.8}_{-3.5}\pm1.5)\times10^{-4}$, where the first uncertainties are statistical and the second systematic.  

\begin{figure}[htbp!]
\centering
{\includegraphics[width=0.8\linewidth]{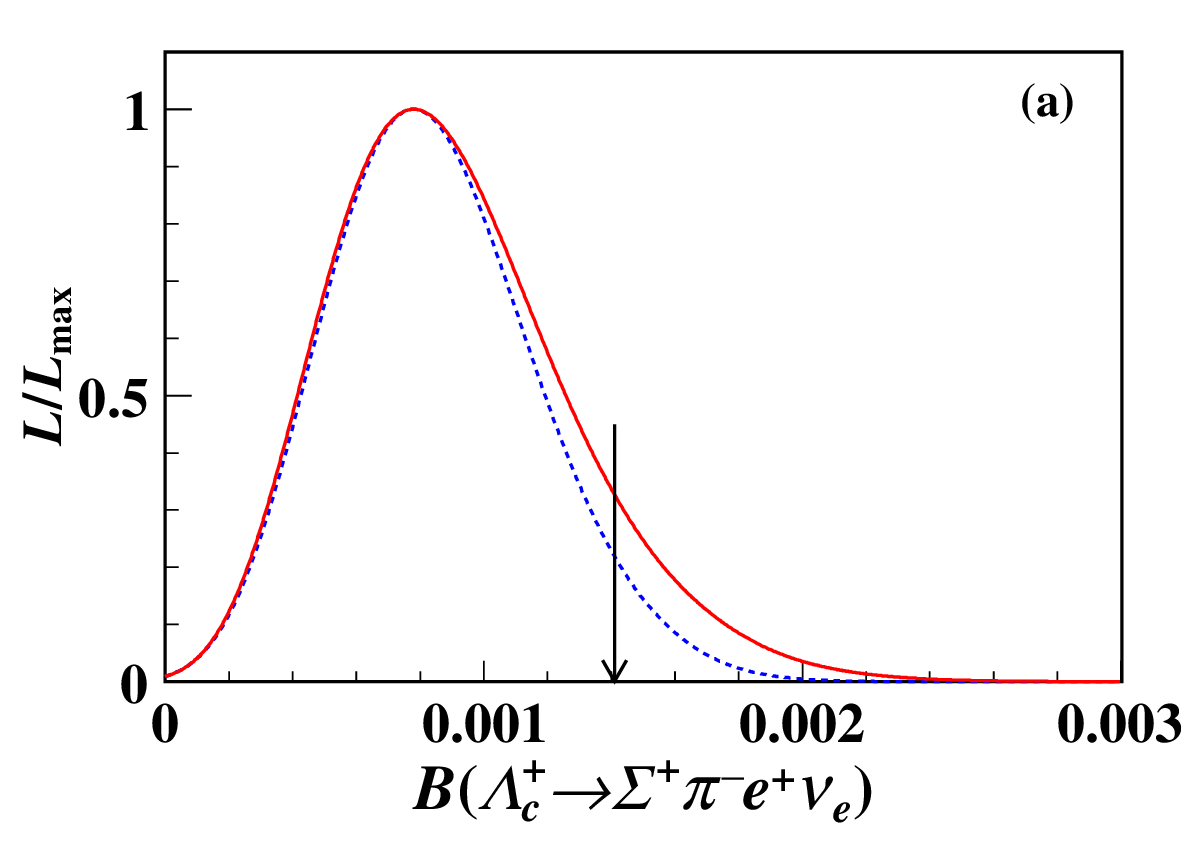}}
{\includegraphics[width=0.8\linewidth]{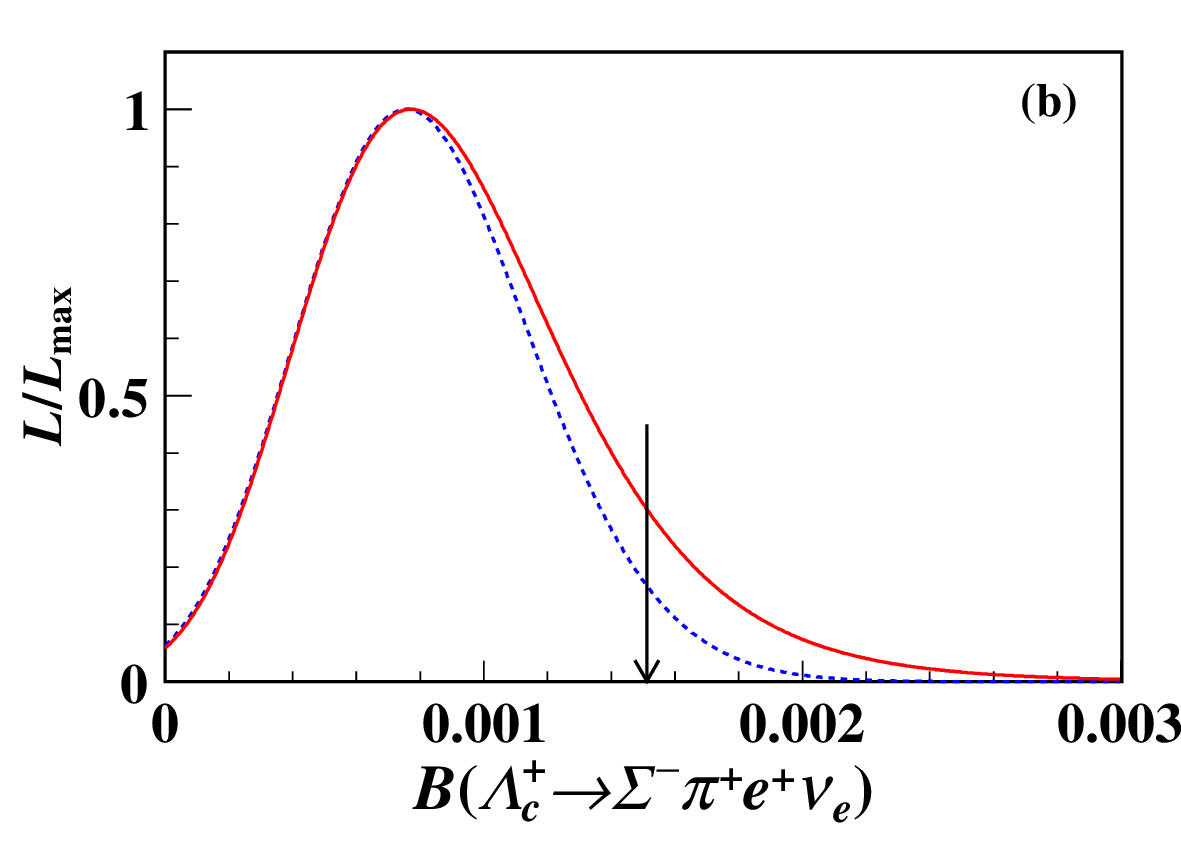}}
\caption{\label{fig:UL}The likelihood curves of (a) ${\mathcal B}(\lambdacp \to \sgmppimev)$ and (b) ${\mathcal B}(\lambdacp \to \sgmmpipev)$.
The red solid and blue dashed lines represent the likelihood curve for the fit, with and without considering the systematic uncertainties. The black arrow indicates the upper limits on BFs at 90\% confidence level. 
}
\end{figure}

Furthermore, as the significances of $\lambdacp \to \sgmppimev$ and $\lambdacp \to \sgmmpipev$  are less than 3$\sigma$ in separate fits, we evaluate the upper limits on their BFs at $90\%$ confidence level by calculating the likelihood values for a series of fixed BF values to obtain their likelihood curves. 
The background description that gives the most conservative upper limit value is used to take into account the effect of additive systematic uncertainty.
Then, the upper limits are evaluated by integrating the likelihood curve from zero to $90\%$. 
The likelihood curves for $\lambdacp \to \sgmppimev$ and $\lambdacp \to \sgmmpipev$ considering both the additive and multiplicative systematic uncertainties are shown in Figs.~\ref{fig:UL}(a),~\ref{fig:UL}(b), respectively. 
Their BF upper limits at $90\%$ confidence level are $\mathcal{B}(\lambdacp \to \sgmppimev) < 1.41 \times 10^{-3}$ and $\mathcal{B}(\lambdacp \to \sgmmpipev) < 1.51 \times 10^{-3}$.

\section{Summary}

\begin{figure*}[htb!]
\centering
{\includegraphics[width=0.325\linewidth]{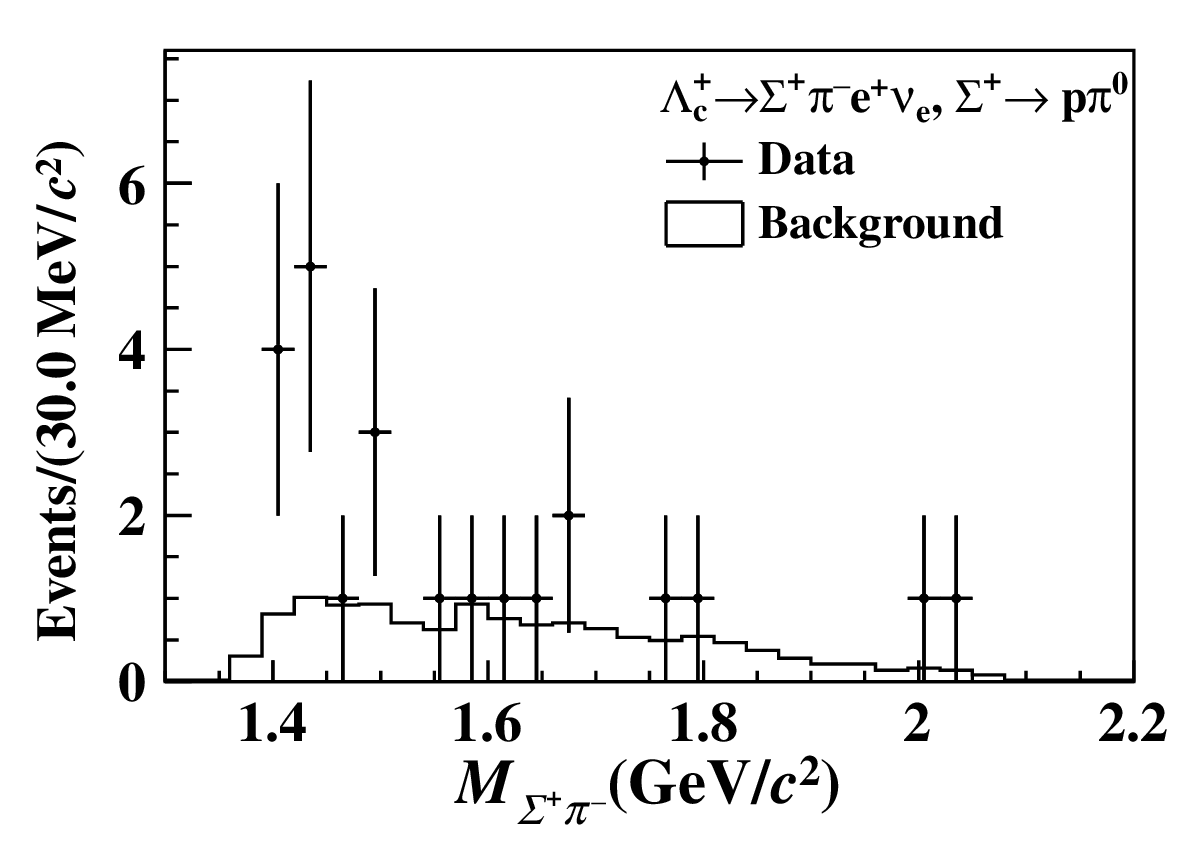}}
{\includegraphics[width=0.325\linewidth]{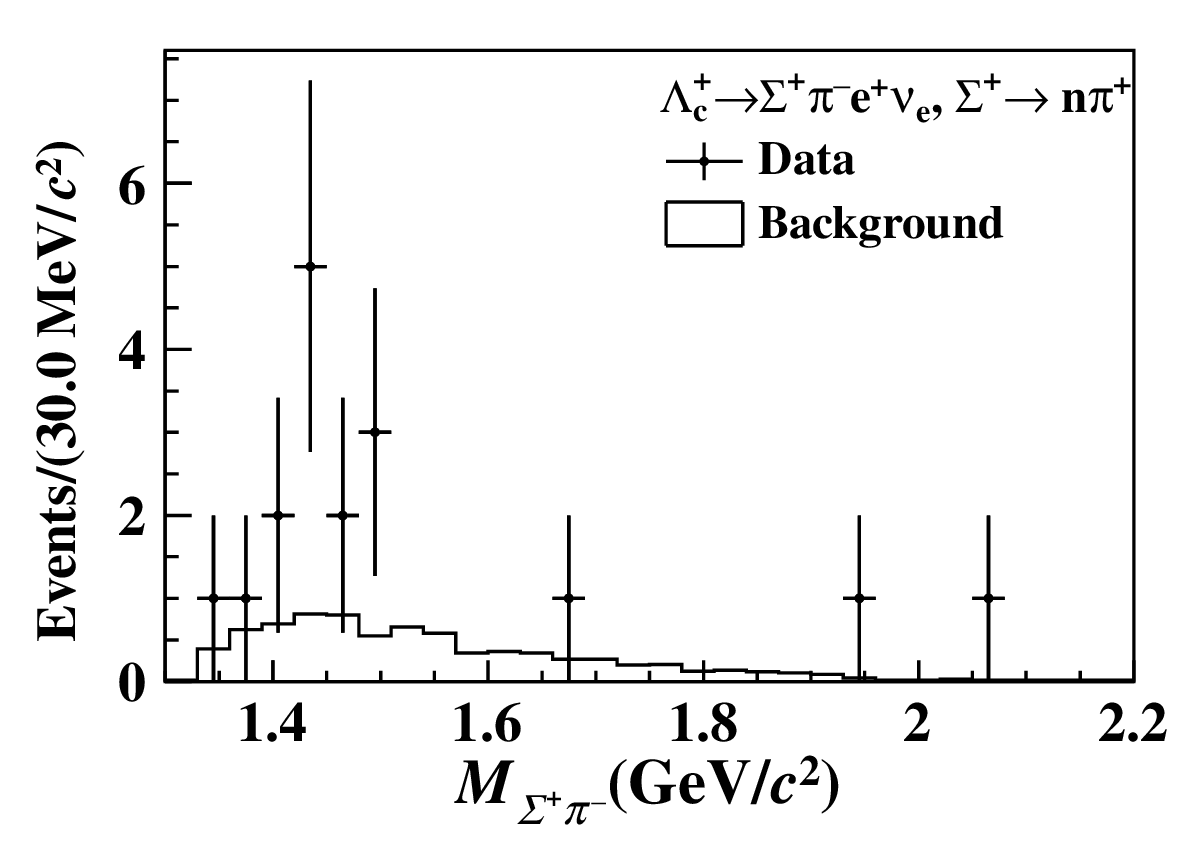}}
{\includegraphics[width=0.325\linewidth]{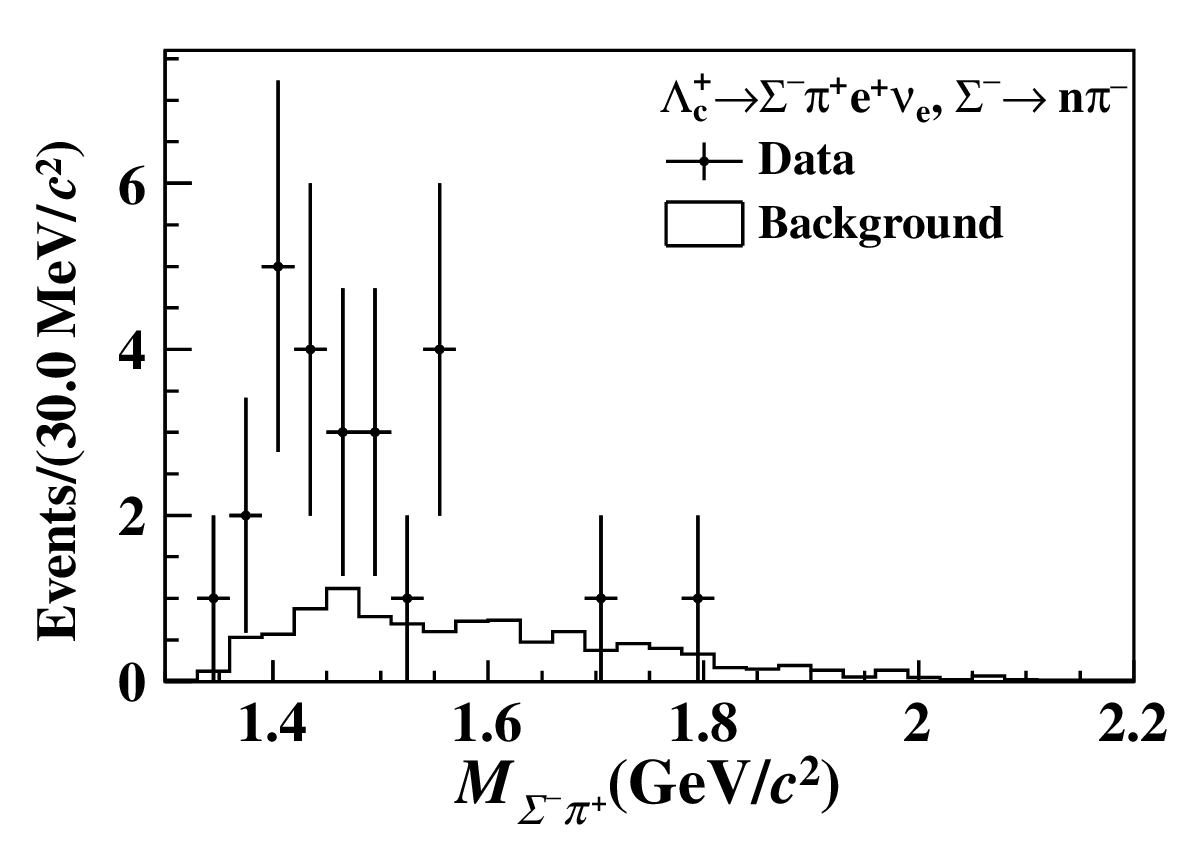}}
\caption{The $M_{\Sigma\pi}$ distribution of (a) $\lambdacp \to \sgmppimev,\, \sgmpToppi$, (b) $\lambdacp \to \sgmppimev,\, \sgmpTonpip$, and (c) $\lambdacp \to \sgmmpipev,\, \sgmmTonpim$ candidates. The black dots with error bars represent data and the histograms represent the MC estimated background. The yields of the MC estimated background are scaled to the numbers from the simultaneous fit results in Fig.~\ref{fig:fit}.
}
\label{fig:Msgmpi}
\end{figure*}

In summary, using $4.5\, \unit{fb^{-1}}$ of $e^+e^-$ collision data collected at center-of-mass energies from 4.600 to 4.699~GeV by the BESIII detector, we present a study of the SL decays $\lambdacp \to \sgmppimev$ and $\lambdacp \to \sgmmpipev$. 
Combining the two SL decay modes by assuming they have equal BF under isospin symmetry, evidence for $\lambdacp \to \Sigma^\pm\pi^\mp e^+\nu_e$ is found with a significance of $3.6\sigma$. The BF is measured to be $\mathcal{B}(\lambdacp \to \Sigma^\pm\pi^\mp e^+\nu_e) = (7.7^{+2.5}_{-2.3}\pm1.3)\times 10^{-4}$.
In the case of separate fits to the two signal modes without considering isospin symmetry, the upper limits of their BFs at 90\% confidence level are $\mathcal{B}(\lambdacp \to \sgmppimev) < 1.41 \times 10^{-3}$ and $\mathcal{B}(\lambdacp \to \sgmmpipev) < 1.51 \times 10^{-3}$. 
Assuming all $\Sigma\pi$ comes from $\Lambda^*$ and isospin symmetry, our results are consistent with the $\mathcal{B}(\lambdacp\to \Lambda^* e^+\nu_e \to \Sigma\pi 
 e^+ \nu_e)$ predictions based on a quark model~\cite{Hussain:2017lir} within two standard deviations. 
The $M_{\Sigma\pi}$ distributions shown in Fig.~\ref{fig:Msgmpi} hint the existence of the $\Lambda^*$ resonance in the $\lambdacp\to\Sigma\pi e^+ \nu_{e}$ process. 
More data samples are anticipated to be collected at the center-of-mass energy near $\lambdacp\lambdacm$ threshold with the BESIII detector in the future~\cite{BESIII:2020nme}. With larger datasets, the SL decays of $\lambdacp$ to $\Lambda(1405)$ and $\Lambda(1520)$ in the $\lambdacp\to\Sigma\pi e^+ \nu_{e}$ process can be studied to better constrain theoretical calculations of $\Lambda_c^+\to\Lambda^*$ form factors.
Furthermore, this will provide a new platform to study the spectroscopy of the excited states of $\Lambda$ baryons.

%%%%%%%%%%%%%%%%%%%%%%%%%%%%%%%%%%%%%%%%%%%%%%%%%%%%%%%%%%%%%%%%
%%%%%     Acknowledgments                          %%%%%%%%%%%%%
%%%%%%%%%%%%%%%%%%%%%%%%%%%%%%%%%%%%%%%%%%%%%%%%%%%%%%%%%%%%%%%%
%% Saved at => 2025-04-10
\acknowledgments
The BESIII Collaboration thanks the staff of BEPCII (https://cstr.cn/31109.02.BEPC) and the IHEP computing center for their strong support. This work is supported in part by National Key R\&D Program of China under Contracts Nos. 2023YFA1606000, 2023YFA1606704; National Natural Science Foundation of China (NSFC) under Contracts Nos. 11635010, 11935015, 11935016, 11935018, 12022510, 12025502, 12035009, 12035013, 12061131003, 12192260, 12192261, 12192262, 12192263, 12192264, 12192265, 12221005, 12225509, 12235017, 12375090, 12361141819; the Chinese Academy of Sciences (CAS) Large-Scale Scientific Facility Program; CAS under Contract No. YSBR-101; 100 Talents Program of CAS; The Institute of Nuclear and Particle Physics (INPAC) and Shanghai Key Laboratory for Particle Physics and Cosmology; ERC under Contract No. 758462; German Research Foundation DFG under Contract No. FOR5327; Istituto Nazionale di Fisica Nucleare, Italy; Knut and Alice Wallenberg Foundation under Contracts Nos. 2021.0174, 2021.0299; Ministry of Development of Turkey under Contract No. DPT2006K-120470; National Research Foundation of Korea under Contract No. NRF-2022R1A2C1092335; National Science and Technology fund of Mongolia; Polish National Science Centre under Contract No. 2024/53/B/ST2/00975; STFC (United Kingdom); Swedish Research Council under Contract No. 2019.04595; U. S. Department of Energy under Contract No. DE-FG02-05ER41374

%\end{linenumbers}

\bibliographystyle{apsrev4-2}
\bibliography{ref}
\end{document}